\pdfoutput=1
\documentclass[a4paper]{llncs}
\usepackage[T1]{fontenc}

\usepackage{amsmath, amssymb}
\usepackage{graphicx}
\usepackage{array}
\usepackage{float}
\usepackage[section]{placeins}
\usepackage[most]{tcolorbox}
\usepackage{enumitem}
\usepackage{cite}
\usepackage{marvosym}
\usepackage[hidelinks]{hyperref}
\usepackage{tikz}
\usetikzlibrary{decorations.pathreplacing}
\usetikzlibrary{calc}

\newif\iflongversion
\longversiontrue
\iflongversion
\newcommand{\longversion}[1]{#1}
\newcommand{\shortversion}[1]{}
\else
\newcommand{\longversion}[1]{}
\newcommand{\shortversion}[1]{#1}
\fi
\newcommand{\fullver}{\longversion{the appendix}\shortversion{the full version~\cite{fullversion}}}

\iflongversion
\usepackage[bibliography=common]{apxproof}
\else
\usepackage[appendix=strip,bibliography=common]{apxproof}
\fi
\newtheoremrep{theorem}{Theorem}
\newtheoremrep{lemma}[theorem]{Lemma}
\newtheoremrep{corollary}[theorem]{Corollary}

\newtcolorbox{problembox}[1]{%
	enhanced, sharp corners, colback=white, colframe=black,
	boxrule=0.8pt,
	left=6pt, right=6pt, top=3pt, bottom=3pt,
	before skip=8pt, after skip=8pt,
	title={#1}, fonttitle=\bfseries, coltitle=black, colbacktitle=black!8}

\tikzset{
	sel/.style  = {circle,draw=black,thick,fill=black,inner sep=0pt,minimum size=4.4pt},
	nsel/.style = {circle,draw=black,thick,fill=white,inner sep=0pt,minimum size=4.4pt},
	ed/.style   = {thin,black!75},
	lab/.style  = {font=\scriptsize,inner sep=1.2pt},
	dts/.style  = {font=\scriptsize,inner sep=0pt}
}

\newcommand{\ud}{\mathrm{ud}}
\newcommand{\UD}{\mathit{UD}}
\title{Full-Cover and Non-Overlapping Uni-domination in Graphs: Bounds and Complexity}
\titlerunning{Full-Cover and Non-Overlapping Uni-domination}

\author{Florent Foucaud\orcidID{0000-0001-8198-693X} \and
	Yan Gerard\orcidID{0000-0002-2664-0650} \and
	Christian Laforest\orcidID{0009-0004-4027-0117} \and
	Le~Vu~Pham\textsuperscript{(\Letter)}\orcidID{0009-0009-6873-8363}}
\authorrunning{F. Foucaud, Y. Gerard, C. Laforest, L. V. Pham}
\institute{CNRS, Clermont Auvergne INP, Mines Saint-\'Etienne, \\
	LIMOS, Clermont-Ferrand, 63000, France\\
	\email{florent.foucaud@uca.fr}, \email{yan.gerard@uca.fr},\\
	\email{christian.laforest@isima.fr}, \email{levupham96@gmail.com}
}

\begin{document}
	\longversion{\pagestyle{plain}}
	\maketitle
	\begin{abstract}
		A vertex is uni-dominated by a set $S \subseteq V(G)$ if exactly one vertex of $S$ lies in its closed neighborhood, and the uni-domination number $\ud^*(G)$ is the largest number of vertices a single set can uni-dominate. The graphs admitting an efficient dominating set are those with $\ud^*(G) = n$. Hedetniemi and Rall asked for the largest constant $c$ with $\ud^*(T) \geq c \cdot n$ for every tree $T$ of order $n$. We answer their question: the constant is $4/5$, and a set attaining it can be computed in linear time. This set is moreover a dominating set. Recall that an efficient dominating set satisfies two requirements at once: it dominates every vertex, and it dominates no vertex twice. Our set satisfies the first one, which leads us to study $\ud^*$ under each of the two requirements separately, giving the Full-Cover variant $\ud^*_0$ and the Non-Overlapping variant $\ud^*_+$. The two behave very differently: on arbitrarily large graphs $\ud^*_0$ may remain bounded by a constant, whereas $\ud^*_+$ always grows with the order. We present bounds for both, and show that the associated decision problems are polynomial for constant thresholds and NP-complete for the thresholds $\lceil \sqrt[c]{n} \rceil$ and $4\sqrt{n}$, respectively, the latter even on split graphs. Finally, we show that neither of the two parameters bounds the other.
	\end{abstract}
	\section{Introduction}\label{sec:intro}
	All graphs here are finite, simple and undirected. For a vertex $v$ of $G = (V,E)$, $N(v)$ is its \emph{open neighborhood} and $N[v] = N(v) \cup \{v\}$ its \emph{closed neighborhood}; $v$ is \emph{dominated} by $S \subseteq V$ if $|N[v] \cap S| \ge 1$, and $S$ is a \emph{dominating set} if it dominates every vertex \cite{ore1962theory,haynes1998fundamentals,haynes2023domination}. Requiring every vertex to be dominated \emph{exactly once} is much stronger: such a set is an \emph{efficient dominating set} \cite{bange1988efficient}, or \emph{perfect code} \cite{biggs1973perfect}, and the closed neighborhoods of its vertices partition $V$. A graph admitting such a set is called \emph{efficient}. Such sets need not exist ($C_5$ admits none), and recognizing the graphs that admit one is NP-complete \cite{bange1988efficient,milanic2013complexity}.

	A closely related notion is that of a \emph{private neighbor} \cite{cockayne1978properties,bollobas1979graph}: for $U \subseteq V$, a vertex $w$ is an \emph{external} private neighbor with respect to $U$ if $w \notin U$ and $w$ has exactly one neighbor in $U$, and a \emph{self} private neighbor if $w \in U$ and $N(w) \cap U = \emptyset$. Hedetniemi and Rall \cite{HedetniemiRall2026} shifted the objective from the sets $U$ to the private neighbors they produce, defining seven maximization parameters based on different types of private neighbors. One of them, $\mathrm{ESPN}(G)$, maximizes the total number of external and self private neighbors over all $U \subseteq V$, as $w$ is of one of these two types if and only if $|N[w] \cap U| = 1$, it is the maximum number of vertices a single set can dominate exactly once. They study it structurally and leave its complexity open.

	We study this quantity as an optimization problem, under the domination-oriented name of \emph{uni-domination}. A vertex $v$ is \emph{uni-dominated} by $S \subseteq V$ if $|N[v] \cap S| = 1$, we write $\UD(S) = \{v : |N[v] \cap S| = 1\}$, $\ud(S) = |\UD(S)|$, and $\ud^*(G) = \max\{\ud(S) : S \subseteq V\}$, the \emph{uni-domination number} of $G$.

	Thus $\ud^*(G) = \mathrm{ESPN}(G)$, and $G$ is efficient if and only if $\ud^*(G) = |V|$. A set realizing $\ud^*(G)$ may leave vertices \emph{undominated} and may dominate others \emph{more than once}, and the two defects are not equally acceptable in applications: for facilities serving a whole network, leaving a vertex uncovered is forbidden while serving one twice only wastes resources, for transmitters whose signals interfere as soon as two of them reach a common vertex, overlaps are forbidden while an uncovered vertex is merely not served. Accordingly we write
	$V_0(S)$, $\UD(S)$ and $V_+(S)$ for the sets of vertices $u$ with $|N[u] \cap S|$ equal to $0$, equal to $1$ and greater than $1$, the \emph{undominated}, \emph{uni-dominated} and \emph{over-dominated} vertices, respectively. Forbidding one defect while maximizing the number of uni-dominated vertices gives the two parameters studied here: the \emph{Full-Cover variant}, requiring $S$ to be a dominating set, and the \emph{Non-Overlapping variant}, requiring the closed neighborhoods of the vertices of $S$ to be pairwise disjoint, respectively,
	$$ \ud_0^*(G) = \max_{S \subseteq V} \{ \ud(S) : V_0(S) = \emptyset \}, \qquad \ud_+^*(G) = \max_{S \subseteq V} \{ \ud(S) : V_+(S) = \emptyset \}. $$
	Each relaxes efficient domination in one direction: $G$ of order $n$ is efficient if and only if $\ud_0^*(G) = n$, if and only if $\ud_+^*(G) = n$, and both are bounded above by the unrestricted parameter, $\max\{\ud_0^*(G), \ud_+^*(G)\} \le \ud^*(G)$. Their behaviors turn out to be of a different nature. The second one places our study inside an existing line of work: it is the efficient domination number of Bange, Barkauskas and Slater \cite{bange1988efficient}, as recalled below, and our contribution to it is the hardness result of Section~\ref{sec:no-complexity} together with the comparison with $\ud_0^*$ of Section~\ref{sec:compare}.
	\subsection{Preliminaries}\label{sec:prelim}
	Throughout, $G = (V,E)$ is a finite simple undirected graph of order $n = |V|$, $\deg(u) = |N(u)|$ for any vertex $u \in V$, and $\Delta$ is the maximum degree of the vertices of $G$. For $X \subseteq V$, the subgraph \emph{induced} by $X$, denoted $G[X]$, has vertex set $X$ and contains exactly the edges of $G$ with both ends in $X$. We write $G - X = G[V \setminus X]$ and $N[X] = \bigcup_{v \in X} N[v]$. A vertex adjacent to all others is \emph{universal}, a vertex $v \neq u$ with $uv \notin E$ is a \emph{non-neighbor} of $u$, and $S \subseteq V$ is a \emph{$2$-packing} \cite{meir1975relations,haynes2023domination} if its vertices are pairwise at distance at least $3$, equivalently if their closed neighborhoods are pairwise disjoint.

	A dominating set is \emph{minimal} if none of its proper subsets is dominating. Following Cockayne, Hedetniemi and Miller \cite{cockayne1978properties}, a set $S$ is \emph{irredundant} if every $v \in S$ has a \emph{private neighbor} with respect to $S$, that is, a vertex $w$ with $N[w] \cap S = \{v\}$, which is an external private neighbor if $w \neq v$ and a self private neighbor if $w = v$. A dominating set is minimal if and only if it is irredundant \cite{cockayne1978properties,haynes2023domination}. Indeed, as every superset of a dominating set is dominating, a dominating set $S$ is minimal if and only if, for every $v \in S$, the set $S \setminus \{v\}$ leaves some vertex $w$ undominated. As $w$ is dominated by $S$, this means exactly that $N[w] \cap S = \{v\}$, that is, $w$ is a private neighbor of $v$. For all other standard notions we follow \cite{haynes1998fundamentals,haynes2023domination}.

	Both of our reductions start from the following NP-complete problem \cite{gonzalez1985clustering}.
	\begin{problembox}{Restricted Exact Cover by 3-Sets (RX3C)}
		\textbf{Instance:} A finite set $X$ with $|X| = 3p$ and a collection $\mathcal{C}$ of $3$-element subsets of $X$ in which every element of $X$ occurs in exactly three subsets. \\
		\textbf{Question:} Is there a subcollection $\mathcal{C}^* \subseteq \mathcal{C}$ that contains every element of $X$ exactly once?
	\end{problembox}
	\subsection{Related Work}\label{sec:related}
	Closest to this paper is a line of research measuring how far a graph is from being efficient. Bange, Barkauskas and Slater introduced the \emph{efficient domination number} $F(G)$, the maximum number of vertices that can be dominated by a set dominating no vertex more than once, so that $G$ is efficient if and only if ${F(G) = n}$ \cite{bange1988efficient,rubalcaba2007efficient}. In our notation $\ud_+^*(G) = F(G)$, we keep the notation $\ud_+^*$ since our aim is to compare the two constrained variants of $\ud^*$ within one framework. Grinstead and Slater defined the \emph{influence} ${I(S) = \sum_{v \in S}(1 + \deg(v))}$ of a set $S$ and observed that a set dominates no vertex more than once precisely when it is a $2$-packing, whence $\ud_+^*(G) = \max\{I(S) : S \text{ a } 2\text{-packing}\}$ \cite{grinstead1990fractional,rubalcaba2007efficient}. Goddard, Oellermann, Slater and Swart \cite{goddard2000bounds} proved the sharp bounds $\ud_+^*(G) \ge 1 + \sqrt{n-1}$ for every graph and $\ud_+^*(T) \ge \sqrt{8(n+2)} - 4$ for every tree, and deciding whether $\ud_+^*(G) \ge k$ for an input $k$ is NP-complete \cite{bange1988efficient,goddard2000bounds}, whereas $\ud_+^*(T)$ is computable in polynomial time on trees \cite{bange1988efficient}. The companion of $\ud_+^*$ on the domination side is the \emph{redundance} $R(G) = \min\{I(S) : S \text{ dominating}\}$ \cite{grinstead1990fractional}, which differs from $\ud_0^*$: since $I(S) = \sum_{u \in V} |N[u] \cap S|$, the quantity $R(G)$ minimizes the total excess of domination, whereas $\ud_0^*$ maximizes the number of vertices whose excess is zero, and an excess of a given total size may be concentrated on one vertex or spread over many. Excess domination in this sense was studied in \cite{harary1998efficient}, and efficient versions of multiple domination in \cite{rubalcaba2007efficient}.

	Hedetniemi and Rall \cite{HedetniemiRall2026} initiated the study of seven private neighbor parameters, determining $\ud^*/\mathrm{ESPN}$ for paths, cycles and complete bipartite graphs. They also observed that $\ud^*(T) \ge n/2$ for every tree and exhibited a family attaining $4n/5$, which led them to ask for the largest constant $c$ such that $\ud^*(T) \ge c \cdot n$ holds for every tree. We settle this question.
	\subsection{Our Results}
	\begin{table}[!ht]
		\centering
		\caption{Summary of the results (entries with a reference are not proved here).}
		\label{tab:results}
		\small
		\setlength{\tabcolsep}{3pt}
		\renewcommand{\arraystretch}{1}
		\begin{tabular}{p{1.9cm}p{4.75cm}p{4.75cm}}
			\hline
			& \textbf{Full-Cover} $\ud_0^*(G)$ & \textbf{Non-Overlapping} $\ud_+^*(G)$ \\
			\hline
			Earlier work & introduced here & $= F(G)$ \cite{bange1988efficient} \\
			Lower bound & $\ge 2$ ($n \ge 2$), best possible & $\ge 1 + \sqrt{n-1}$, best possible \cite{goddard2000bounds} \\
			Trees & $\ge 4n/5$, best possible & $\ge \sqrt{8(n+2)}-4$, best possible \cite{goddard2000bounds} \\
			Polynomial & $k = c$ constant & $k = c$ constant \\
			NP-complete & $k = \lceil\sqrt[c]{n}\rceil$, $c \ge 2$ fixed & $k = 4\sqrt{n}$, split graphs \\
			\hline
		\end{tabular}
	\end{table}
	Our main result is that every tree $T$ of order $n$ satisfies $\ud_0^*(T) \ge 4n/5$ and that a dominating set attaining this bound can be computed in linear time (Theorem~\ref{thm:tree_full_cover}, Corollaries~\ref{cor:tree_espn} and~\ref{cor:tree_algorithm}). Moreover, as shown in \cite{HedetniemiRall2026}, the constant $4/5$ cannot be improved, so that our result answers Problem~1 of \cite{HedetniemiRall2026}. On general graphs the same variant behaves very differently: we prove constructively that $\ud_0^*(G) \ge 2$ for every graph of order $n \ge 2$ (Theorem~\ref{thm:lower_bound}), and we exhibit, for every constant $c \ge 2$, arbitrarily large graphs with $\ud_0^*(G) = c$ (Theorem~\ref{thm:general_tightness}). The case $c = 2$ shows that the constant $2$ cannot be improved. Deciding whether $\ud_0^*(G) \ge c$ is polynomial for every fixed $c$ (Theorem~\ref{thm:poly_time}), whereas the threshold $\lceil \sqrt[c]{n} \rceil$ makes the problem NP-complete for every fixed $c \ge 2$ (Theorem~\ref{thm:np_complete}).

	For the Non-Overlapping parameter, \cite{goddard2000bounds} proved the sharp bound $\ud_+^*(G) \ge 1 + \sqrt{n-1}$. We observe that their argument applies to every maximal $2$-packing containing a vertex of maximum degree, so that a greedy polynomial-time construction attains the bound, and we show that deciding whether $\ud_+^*(G) \ge 4\sqrt{n}$, a threshold within a constant factor of that bound, is NP-complete even for split graphs (Theorem~\ref{thm:np_complete_sqrt}). Finally, we show that neither of the two constrained parameters bounds the other: there are graphs on which $\ud_+^*$ exceeds $\ud_0^*$ by an arbitrarily large factor, and trees on which the opposite holds (Theorems~\ref{thm:separation_1} and~\ref{thm:separation_2}). Table~\ref{tab:results} summarizes the results. Proofs of statements marked with $\star$ are deferred to \fullver, except for the constructions of the two reductions.
	\section{The Full-Cover Uni-domination Problem}
	Throughout this section $S$ is required to dominate every vertex, that is, $V_0(S) = \emptyset$, and the parameter is $\ud_0^*(G)$.
	\subsection{Structural Properties}
	Our first bound rests on the classical characterization of minimal dominating sets by private neighbors.
	\begin{theoremrep}[$\star$] \label{thm:lower_bound}
		Every graph $G = (V,E)$ with $|V| \ge 2$ satisfies $\ud_0^*(G) \ge 2$, and a dominating set $S$ with $\ud(S) \ge 2$ can be constructed in polynomial time.
	\end{theoremrep}
	\begin{proof}
		We first observe that every minimal dominating set $S$ of $G$ satisfies $\ud(S) \ge |S|$. Indeed, a dominating set is minimal if and only if it is irredundant \cite{cockayne1978properties,haynes2023domination}, so every $v \in S$ has a private neighbor $p_v$ with respect to $S$, that is, a vertex with $N[p_v] \cap S = \{v\}$. In particular $p_v \in \UD(S)$. If $p_u = p_v$ for two distinct $u, v \in S$, then $\{u\} = N[p_u] \cap S = N[p_v] \cap S = \{v\}$, a contradiction, so the mapping $v \mapsto p_v$ is injective and $\ud(S) \ge |S|$.

		Now assume $|V| \ge 2$. A minimal dominating set is obtained in polynomial time by a standard greedy reduction: initialize $S = V$ and remove a vertex $v$ from $S$ whenever $S \setminus \{v\}$ still dominates $G$. One pass over the vertices suffices, and the resulting set $S$ is a minimal dominating set, so $\ud(S) \ge |S|$ by the observation above. If $|S| \ge 2$, we obtain $\ud(S) \ge 2$. If $|S| = 1$, say $S = \{u\}$, then $u$ dominates $G$ on its own and is therefore a universal vertex, so $\UD(S) = V$ and $\ud(S) = |V| \ge 2$. In both cases the polynomially constructed dominating set $S$ is a witness for $\ud_0^*(G) \ge 2$. \qed
	\end{proof}
	Theorem~\ref{thm:lower_bound} raises the question of the tightness of the constant $2$. One might expect the number of uni-dominated vertices to grow with $|V|$, we show that under full coverage it can remain bounded by a constant.

	\begin{toappendix}
		\begin{lemma} \label{lem:non_neighbors}
			Let $G = (V,E)$ be a graph and let $u \in V$ have exactly $k$ non-neighbors in $G$. For any subset $S \subseteq V$ with $|S| \ge k + 2$, the vertex $u$ is not uni-dominated by $S$.
		\end{lemma}
		\begin{proof}
			Write $\overline{N}(u) = V \setminus N[u]$ for the set of the non-neighbors of $u$, so that $|\overline{N}(u)| = k$. Since $V$ is the disjoint union of $N[u]$ and $\overline{N}(u)$, the number of times $u$ is dominated by $S$ equals $|S \cap N[u]| = |S \setminus \overline{N}(u)| \ge |S| - k \ge 2$. Hence $u \in V_+(S)$ and $u$ is not uni-dominated. \qed
		\end{proof}
	\end{toappendix}
	\begin{theoremrep}[$\star$] \label{thm:general_tightness}
		For any integers $c \ge 2$ and $p \ge 1$, there exists a graph $G = (V,E)$ of order $n = c + 2p$ with $\ud_0^*(G) = c$.
	\end{theoremrep}
	\begin{proofsketch}
		Given $c \ge 2$ and $p \ge 1$, we construct $G = (V,E)$ as follows. Let the vertex set $V$ be partitioned into two disjoint subsets: $\mathcal{C} = \{x_1, \dots, x_c\}$ and $R = \{u_1, v_1, \dots, u_p, v_p\}$. We define the edge set $E$ such that there are no edges within $\mathcal{C}$, every vertex in $\mathcal{C}$ is adjacent to every vertex in $R$, and the induced subgraph on $R$ is a complete graph minus a perfect matching $M = \{\{u_i, v_i\} \mid 1 \le i \le p\}$. Formally, ${E = \{xy : x \in \mathcal{C},\, y \in R\} \cup \bigl(\{xy : x, y \in R,\, x \neq y\} \setminus M\bigr)}$. Clearly, $n = |V| = c + 2p$. The construction is illustrated in Figure~\ref{fig:graph_G_horizontal}.
		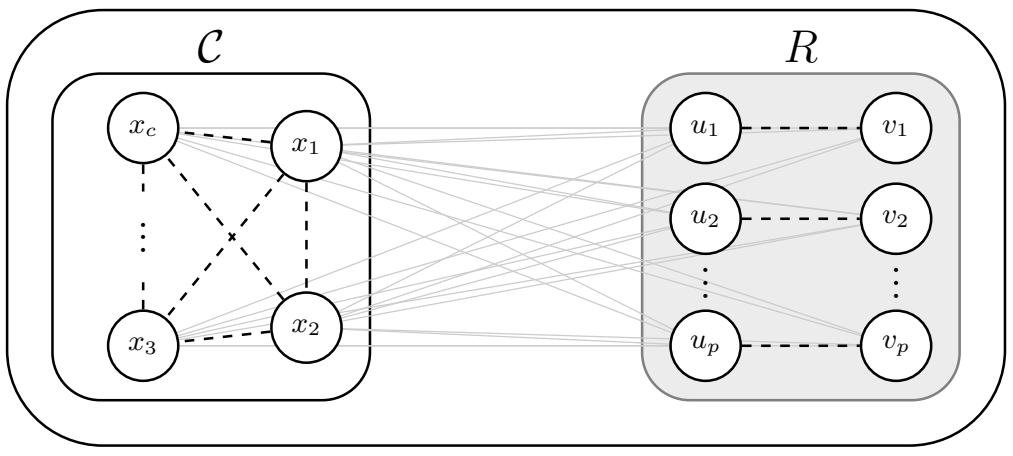
\begin{figure}[!ht]
			\centering
			\scalebox{0.6}{
				\begin{tikzpicture}[
					vertex/.style={circle, draw, thick, fill=white, minimum size=22pt, inner sep=0pt},
					dots/.style={circle, draw=none, minimum size=0pt, inner sep=2pt},
					anti edge/.style={dashed, thick},
					cross edge/.style={gray!40, thin}
					]
					\draw[thick, rounded corners=30pt] (-5.5, -2.3) rectangle (5.5, 2.5);
					\draw[rounded corners=15pt, thick] (-5, -1.8) rectangle (-1.5, 1.8);
					\node at (-3.25, 2.1) {\Large $\mathcal{C}$};
					\draw[fill=gray!15, rounded corners=15pt, thick, draw=black!50] (1.5, -1.8) rectangle (5, 1.8);
					\node at (3.25, 2.1) {\Large $R$};

					\coordinate (X1) at (-2.2, 1);
					\coordinate (X2) at (-2.2, -1);
					\coordinate (X3) at (-4, -1.2);
					\coordinate (XC) at (-4, 1.2);
					\coordinate (U1) at (2.2, 1.2);
					\coordinate (V1) at (4.3, 1.2);
					\coordinate (U2) at (2.2, 0.2);
					\coordinate (V2) at (4.3, 0.2);
					\coordinate (UP) at (2.2, -1.2);
					\coordinate (VP) at (4.3, -1.2);

					\foreach \x in {X1, X2, X3, XC} {
						\foreach \y in {U1, U2, UP, V1, V2, VP} {
							\draw[cross edge] (\x) -- (\y);
						}
					}

					\node[vertex] (x1) at (X1) {$x_1$};
					\node[vertex] (x2) at (X2) {$x_2$};
					\node[vertex] (x3) at (X3) {$x_3$};
					\node[vertex] (xc) at (XC) {$x_c$};
					\node[dots] at (-4, 0.1) {\Large $\vdots$};
					\draw[anti edge] (x1) -- (x2);
					\draw[anti edge] (x2) -- (x3);
					\draw[anti edge] (x1) -- (xc);
					\draw[anti edge] (x1) -- (x3);
					\draw[anti edge] (x2) -- (xc);
					\draw[anti edge] (x3) -- (-4, -0.5);
					\draw[anti edge] (xc) -- (-4, 0.5);

					\node[vertex] (u1) at (U1) {$u_1$};
					\node[vertex] (u2) at (U2) {$u_2$};
					\node[dots] at (2.2, -0.4) {\Large $\vdots$};
					\node[vertex] (up) at (UP) {$u_p$};
					\node[vertex] (v1) at (V1) {$v_1$};
					\node[vertex] (v2) at (V2) {$v_2$};
					\node[dots] (vdots) at (4.3, -0.4) {\Large $\vdots$};
					\node[vertex] (vp) at (VP) {$v_p$};
					\draw[anti edge] (u1) -- (v1);
					\draw[anti edge] (u2) -- (v2);
					\draw[anti edge] (up) -- (vp);
				\end{tikzpicture}
			}
			\caption{Construction of $G$ such that $\ud_0^*(G) = c$. Gray lines are edges and dashed lines are non-edges; the edges inside the shaded set $R$ are omitted.}
			\label{fig:graph_G_horizontal}
		\end{figure}

		The value $c$ is attained, for instance, by $S = \mathcal{C}$: every vertex of $R$ is dominated by all $c \ge 2$ vertices of $\mathcal{C}$, while each vertex of the independent set $\mathcal{C}$ is dominated by itself only, so that $\UD(S) = \mathcal{C}$. Any set consisting of one vertex of $\mathcal{C}$ and one vertex of $R$ is another optimal solution. For the upper bound, let $S$ be a dominating set. As $G$ has no universal vertex, $|S| \ge 2$. Every vertex of $R$ has exactly one non-neighbor, and is therefore dominated at least $|S| - 1$ times. Hence, if $|S| \ge 3$, no vertex of $R$ is uni-dominated and $\ud(S) \le |\mathcal{C}| = c$. The dominating sets of size $2$ are handled by a case analysis given in \fullver. \qed
	\end{proofsketch}
	\begin{proof}
		Let $G$ be the graph constructed in the main text and illustrated in Figure~\ref{fig:graph_G_horizontal}, its order is $c + 2p$. By construction, there are no edges between any two vertices in $\mathcal{C}$, making $\mathcal{C}$ an independent set of size $c$. Every vertex in $\mathcal{C}$ is non-adjacent only to the other $c - 1$ vertices within $\mathcal{C}$. Meanwhile, every vertex in $R$ is connected to all vertices in $\mathcal{C}$ and all vertices in $R$ except its unique non-neighbor. Thus, every vertex in $R$ has exactly one non-neighbor in $G$.

		Let $S$ be a dominating set of $G$. Since $G$ has no universal vertex, we must have $|S| \ge 2$. We evaluate $\ud(S)$ based on the cardinality of $S$:
		\begin{itemize}
			\item \textbf{Case 1:} $|S| \ge 3$.
			For any vertex $v \in R$, its number of non-neighbors is exactly one. By Lemma~\ref{lem:non_neighbors} applied with $k = 1$, since $|S| \ge 3$, no vertex in $R$ can be uni-dominated. Consequently, $\UD(S) \subseteq \mathcal{C}$. Since $|\mathcal{C}| = c$, it follows that $\ud(S) = |\UD(S)| \le c$.
			\item \textbf{Case 2:} $|S| = 2$. Let $S = \{y, z\}$.
			We analyze the possible configurations of $S$:
			\begin{itemize}
				\item If $y, z \in \mathcal{C}$: since $\mathcal{C}$ is an independent set, neither $y$ nor $z$ dominates the other, and no vertex of $\mathcal{C} \setminus \{y, z\}$ is adjacent to either of them. Two sub-cases arise. If $c > 2$, then any $x \in \mathcal{C} \setminus \{y, z\}$ satisfies $x \in V_0(S)$, so $S$ is not a dominating set and this configuration does not occur. If $c = 2$, then $S = \mathcal{C}$: every vertex of $R$ is adjacent to both $y$ and $z$, hence is dominated twice, while $y$ and $z$ are each dominated by themselves only. Consequently $\UD(S) = \mathcal{C}$ and $\ud(S) = 2 = c$.
				\item If $y \in \mathcal{C}$ and $z \in R$: Since $y$ and $z$ are adjacent, they dominate each other, meaning $y, z \in V_+(S)$. For the remaining vertices, any vertex in $\mathcal{C} \setminus \{y\}$ is adjacent to $z$ but not $y$, so it is dominated exactly once. For vertices in $R \setminus \{z\}$, they are all adjacent to $y$. They are also adjacent to $z$, except for the unique non-neighbor $z'$ of $z$. Thus, $z'$ is only dominated by $y$, so $z' \in \UD(S)$. This yields $\UD(S) = (\mathcal{C} \setminus \{y\}) \cup \{z'\}$, giving exactly $\ud(S) = (c - 1) + 1 = c$.
				\item If $y, z \in R$: First, since every vertex in $R$ is adjacent to all vertices in $\mathcal{C}$, $S$ covers $\mathcal{C}$. Second, within $R$, every vertex has exactly one non-neighbor. Since $S$ contains two distinct vertices $y$ and $z$, the only vertex not dominated by $y$ is its unique non-neighbor, which is necessarily dominated by $z$, and vice versa. Thus, the entire graph is dominated ($V_0(S) = \emptyset$). Furthermore, because both $y$ and $z$ are adjacent to all of $\mathcal{C}$, every vertex in $\mathcal{C}$ is dominated exactly twice, meaning $\UD(S) \subseteq R$. Since the non-edges of $R$ form the perfect matching $M$, $\UD(S)$ will contain exactly two vertices (either $y$ and $z$ if $yz \notin E$, or their respective unique non-neighbors if $yz \in E$). Thus, $\ud(S) = 2 \le c$.
			\end{itemize}
			Therefore, for any dominating set of size 2, we have $\ud(S) \le c$.
		\end{itemize}
		From both cases, we establish that $\ud(S) \le c$ for any dominating set $S$.

		This upper bound is achievable through two distinct structural configurations. As shown in Case 2, any dominating set consisting of exactly one vertex of $\mathcal{C}$ and one vertex of $R$ yields exactly $\ud(S) = c$. The second one is $S = \mathcal{C}$, already met in Case 2 for $c = 2$ and available for every $c \ge 2$: since $p \ge 1$, every vertex of $R$ is adjacent to all $c \ge 2$ vertices of $\mathcal{C}$, so $R \subseteq V_+(S)$, and since $\mathcal{C}$ is an independent set, each vertex of $\mathcal{C}$ is dominated by itself only. No vertex is left undominated ($V_0(S) = \emptyset$) and $\UD(S) = \mathcal{C}$, which again yields $\ud(S) = c$. Having established both the upper bound and its attainability, we conclude that $\ud_0^*(G) = c$. \qed
	\end{proof}
	\subsection{Trees}
	While Theorem~\ref{thm:general_tightness} shows that $\ud_0^*(G)$ may be bounded by a constant in general graphs, on trees the situation is different: a constant fraction of the vertices can always be uni-dominated by a dominating set.
	\begin{theorem} \label{thm:tree_full_cover}
		Every tree $T$ of order $n$ satisfies $\ud_0^*(T) \ge \dfrac{4n}{5}$.
	\end{theorem}
	\begin{proof}
		We proceed by induction on the order $n$ of the tree. Throughout the proof, the main cases are illustrated in Figure~\ref{fig:tree-cases}.

		\begin{figure}[ht!]
			\begin{center}
				\includegraphics[width=1\textwidth]{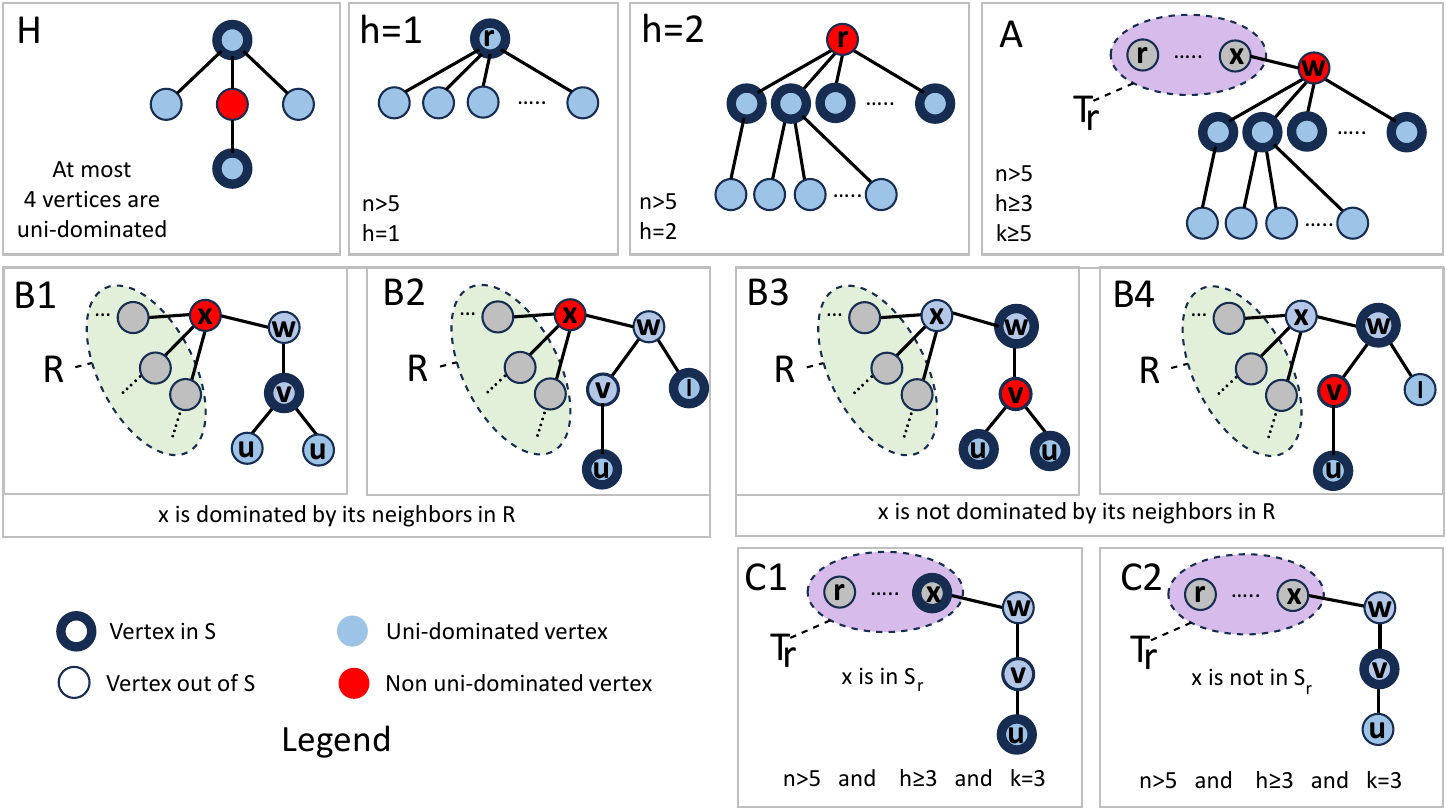}
			\end{center}
			\caption{Overview of the main cases of the proof of Theorem~\ref{thm:tree_full_cover}.}
			\label{fig:tree-cases}
		\end{figure}

		We start with the small cases. The orders $n \le 2$ are immediate. If $n = 3$, the tree is a path and its central vertex uni-dominates all three vertices. If $n = 4$, the tree is a path or a star, and in both cases it admits an efficient dominating set. Suppose now that $n = 5$. There are three non-isomorphic trees of order $5$. If $T$ is a star, its center uni-dominates every vertex. If $T$ is a path, two vertices at distance $3$ uni-dominate every vertex. It remains to consider the tree $H$ obtained from $K_{1,3}$ with center $c$ and leaves $\ell_1, \ell_2, d$ by subdividing the edge $cd$ with a new vertex $e$. The set $S = \{c, d\}$ is dominating and it uni-dominates $c$, $\ell_1$, $\ell_2$ and $d$, while $e$ is dominated twice. Moreover, no subset of $V(H)$ uni-dominates more than $4$ vertices, so the bound $4/5$ is tight. In all these small cases $\ud_0^*(T) \ge 4n/5$.

		Now let $n \ge 6$ and assume that every tree of order less than $n$ satisfies the statement. Root $T$ at an arbitrary vertex $r$ and let $h$ be the height of the resulting rooted tree. If $h \le 1$, then $T$ is a star and $S = \{r\}$ uni-dominates all $n$ vertices. If $h = 2$, let $S$ be the set of children of $r$: every vertex is dominated, each child is uni-dominated by itself since two children of $r$ are never adjacent and no grandchild belongs to $S$, and each grandchild is uni-dominated by its parent. Only $r$ may be dominated more than once, so $\ud(S) \ge n - 1 \ge 4n/5$ since $n \ge 6$. We may therefore assume $h \ge 3$. Let $u$ be a vertex of maximum depth. Since $h \ge 3$, it has a parent $v$, a grandparent $w$ and a great-grandparent $x$. By maximality of the depth, $u$ is a leaf, every child of $v$ is a leaf, and every child of $w$ is either a leaf or a vertex all of whose children are leaves. Let $T_w$ be the subtree rooted at $w$ and put $k = |V(T_w)| \ge 3$. The only edge joining $T_w$ to the rest of $T$ is $wx$. Consequently $T_r = T - V(T_w)$ is again a \emph{tree}, of order $n - k \ge 1$ since it contains $x$. We distinguish three cases according to $k$.

		\noindent \textbf{Case 1:} $k \ge 5$ (panel A of Figure~\ref{fig:tree-cases}). Let $v_1, \dots, v_q$ be the children of $w$, with ${v_1 = v}$. By the choice of $u$, every child of every $v_i$ is a leaf. By the induction hypothesis, $T_r$ has a dominating set $S_r$ with $\ud(S_r) \ge 4(n-k)/5$. Put ${S_w = \{v_1, \dots, v_q\}}$ and $S = S_r \cup S_w$. Each $v_i$ is uni-dominated by itself, since $w \notin S_w$ and none of its children belongs to $S_w$. Each child of $v_i$ is a leaf uni-dominated by $v_i$. The vertex $w$, being adjacent to all the $q \ge 1$ vertices of $S_w$, is dominated but possibly more than once, so we do not count it. Hence $S_w$ dominates $T_w$ and uni-dominates at least $k-1$ of its vertices. The two parts do not interfere: every vertex of $S_w$ has all its neighbors inside $T_w$, so $S_w$ dominates no vertex of $T_r$, and conversely $w$ is the only vertex of $T_w$ with a neighbor outside $T_w$. Therefore $S$ is a dominating set of $T$ and, using $k - 1 \ge 4k/5$, we obtain ${\ud(S) \ge 4(n-k)/5 + (k-1) \ge 4n/5}$.

		\noindent \textbf{Case 2:} $k = 4$ (panels B1--B4 of Figure~\ref{fig:tree-cases}). Exactly two configurations are possible: either $w$ has a single child $v$ and $v$ has exactly two leaf children $u_1, u_2$, or $w$ has exactly two children, namely $v$ with a single leaf child $u$, and a leaf $\ell$. Here we remove $x$ as well: let $R = V(T) \setminus (V(T_w) \cup \{x\})$, so $|R| = n - 5$. Since $x$ is removed, the forest $T[R]$ may be disconnected. Let $T_1, \dots, T_m$ be its components. Each is a tree of order less than $n$, so the induction hypothesis provides a dominating set $S_i$ of $T_i$ with $\ud(S_i) \ge 4|V(T_i)|/5$. Put ${S_0 = S_1 \cup \dots \cup S_m}$. As no edge joins two components, $S_0$ dominates $T[R]$ and uni-dominates at least $4(n-5)/5$ of its vertices. We choose $S_w \subseteq V(T_w)$ according to whether $x$ is dominated by its neighbors in $R$, that is, whether $x$ has a neighbor in $S_0$, and set $S = S_0 \cup S_w$. Since $S_0 \subseteq R$ and $S_w \subseteq V(T_w)$, we have $x \notin S$.

		Suppose first that $x$ is dominated by its neighbors in $R$, as in panels B1 and B2. The vertex $x$ is then dominated, possibly several times, and we do not count it among the uni-dominated vertices. The four vertices of $T_w$, on the other hand, are all uni-dominated. In panel B1 we take $S_w = \{v\}$: the vertex $v$ is uni-dominated by itself, since none of $u_1, u_2, w$ lies in $S_w$, and $u_1$, $u_2$ and $w$ are uni-dominated by $v$, the only other neighbor of $w$ being $x \notin S$. In panel B2 we take $S_w = \{u, \ell\}$: the vertices $u$ and $\ell$ are uni-dominated by themselves, $v$ is uni-dominated by $u$, and $w$ by $\ell$, again because $x \notin S$.

		Suppose now that $x$ is not dominated by its neighbors in $R$, as in panels B3 and B4. We then put $w$ in $S_w$, so that $w$ is uni-dominated by itself, since $x \notin S$ and $v \notin S_w$, and $x$ is uni-dominated by $w$, its only neighbor in $S$. In panel B3 we take $S_w = \{w, u_1, u_2\}$: $u_1$ and $u_2$ are uni-dominated by themselves since $v \notin S_w$, only $v$ is over-dominated, being adjacent to the three vertices of $S_w$. In panel B4 we take $S_w = \{w, u\}$: the vertex $u$ is uni-dominated by itself, $\ell$ is uni-dominated by $w$, only $v$ is over-dominated, being adjacent to both $w$ and $u$.

		In each of the four configurations every vertex of $V(T_w) \cup \{x\}$ is dominated and at least four of these five vertices are uni-dominated. Moreover every vertex of $S_w$ has all its neighbors in $V(T_w) \cup \{x\}$, so $S_w$ dominates no vertex of $R$. Conversely $x$ is the only vertex outside $R$ with a neighbor in $R$, and $x$ is counted only in panels B3 and B4, where it has no neighbor in $S_0$. Hence $S$ is a dominating set of $T$ and $\ud(S) \ge 4(n-5)/5 + 4 = 4n/5$.

		\noindent \textbf{Case 3:} $k = 3$. Here $w$ has a single child $v$, which in turn has a single leaf child $u$. By the induction hypothesis, the tree $T_r$, of order $n-3$, has a dominating set $S_r$ with $\ud(S_r) \ge 4(n-3)/5$. This time the choice of $S_w$ depends on whether $x$ belongs to $S_r$, the two resulting configurations being drawn in panels C1 and C2 of Figure~\ref{fig:tree-cases}. If $x \in S_r$, take $S_w = \{u\}$: then $u$ is uni-dominated by itself, $v$ by $u$ since $w \notin S_w$, and $w$ by $x$ since $v, w \notin S_w$. If $x \notin S_r$, take $S_w = \{v\}$: then $v$ uni-dominates $u$, itself and $w$, the latter because the only other neighbor of $w$ is $x \notin S$. In both cases the three vertices of $T_w$ are uni-dominated, and since $N[u] \subseteq V(T_w)$ and $N[v] \subseteq V(T_w)$, the set $S_w$ dominates no vertex of $T_r$. Hence $S = S_r \cup S_w$ is a dominating set of $T$ and $\ud(S) \ge 4(n-3)/5 + 3 \ge 4n/5$.

		In all cases we have exhibited a dominating set $S$ of $T$ with $\ud(S) \ge 4n/5$, which completes the induction. \qed
	\end{proof}
	Since every dominating set is also an admissible set for unrestricted
	uni-domination, Theorem~\ref{thm:tree_full_cover} immediately yields the same lower bound for
	$\ud^*$. Moreover, the tree $H$ considered in the proof of Theorem~\ref{thm:tree_full_cover}
	shows that the constant cannot be improved, even without the domination
	requirement.
	\begin{corollary} \label{cor:tree_espn}
		Every tree $T$ of order $n$ satisfies
		$\ud^*(T)\ge \ud_0^*(T)\ge \dfrac{4n}{5}$.
		Moreover, the constant $4/5$ is best possible for both parameters.
	\end{corollary}
	\begin{proof}
		The inequalities follow from Theorem~\ref{thm:tree_full_cover} and
		$\ud_0^*(T)\le \ud^*(T)$. For sharpness, consider the tree $H$ in Figure~\ref{fig:tree-cases}. It has no efficient dominating set. \qed
	\end{proof}
	Corollary~\ref{cor:tree_espn} answers Problem~1 of Hedetniemi and Rall~\cite{HedetniemiRall2026}. We also note that~\cite{HedetniemiRall2026} provides arbitrarily large trees for which no more than $4/5$ of the vertices can be uni-dominated. Moreover, the proof of
	Theorem~\ref{thm:tree_full_cover} is constructive, and its local reductions can be organized bottom-up so that every vertex and edge is processed a constant number of times.
	\begin{corollaryrep}[$\star$] \label{cor:tree_algorithm}
		For every tree $T$ of order $n$, a dominating set $S$ of $T$ with $\ud(S) \ge \dfrac{4n}{5}$ can be computed in $\mathcal{O}(n)$ time.
	\end{corollaryrep}
	\begin{proof}
		We turn the induction of Theorem~\ref{thm:tree_full_cover} into a bottom-up peeling procedure. The difference with a direct recursive implementation is that the tree is rooted only once and that no traversal is recomputed after a deletion.

		\textbf{Initialization.} Root $T$ at an arbitrary vertex $r$. One traversal computes, for every vertex, its parent, its list of children and its depth, and groups the vertices into buckets $D_0, D_1, \dots$ by depth. Every vertex is marked \emph{active}, and we maintain the number of active vertices. Depths are never recomputed: they are those of the initial rooting.

		\textbf{Invariant.} The active vertices always induce a subtree of $T$ containing $r$, which we call the \emph{active tree}, because each step deletes a whole rooted subtree of it. Consequently the depth of an active vertex in the active tree is its depth in $T$, and the height of the active tree is the largest depth of an active vertex.

		\textbf{One step.} Let $u$ be an active vertex of maximum depth. If at most $5$ vertices are active, or if $\mathrm{depth}(u) \le 2$, the active tree is \emph{terminal} and is treated as described below. Otherwise $u$ has a parent $v$, a grandparent $w$ and a great-grandparent $x$, all active. We let $T_w$ be the active subtree rooted at $w$ and $k = |V(T_w)|$. Deleted vertices are grouped into \emph{bags}: some bags are determined at once, the others store a rule depending on the membership in $S$ of a single active vertex, to which the bag is then \emph{attached}. The three cases of Theorem~\ref{thm:tree_full_cover} are handled as follows.

		\emph{Case $k \ge 5$.} The bag is $V(T_w)$ and the choice $S_w = \{v_1, \dots, v_q\}$, where $v_1, \dots, v_q$ are the active children of $w$, is fixed immediately. By Case 1 of Theorem~\ref{thm:tree_full_cover} every vertex of the bag is dominated, at least $k - 1 \ge 4k/5$ of them are uni-dominated, and no vertex of $S_w$ has a neighbor outside the bag.

		\emph{Case $k = 3$.} The bag is $V(T_w) = \{w, v, u\}$, it is attached to $x$ and stores the rule of Case 3 of Theorem~\ref{thm:tree_full_cover}, namely $S_w = \{u\}$ if $x \in S$ and $S_w = \{v\}$ if $x \notin S$. Under both alternatives $w \notin S_w$, so no vertex of the bag dominates $x$, and all three vertices of the bag are uni-dominated.

		\emph{Case $k = 4$.} Here $Q = V(T_w) \cup \{x\}$ has order $5$, and we delete the whole active subtree rooted at $x$, that is, $Q$ together with the subtrees rooted at the active children of $x$ other than $w$, we call the latter the \emph{side components}. Since $u$ has maximum depth, every active vertex has depth at most $\mathrm{depth}(x) + 3$, so a side component, whose root has depth $\mathrm{depth}(x) + 1$, has height at most $2$. Each side component is therefore solved at once: if its order is at most $5$ we use a base case of Theorem~\ref{thm:tree_full_cover}, otherwise we select its root if its height is at most $1$, and the children of its root if its height is $2$. In every case the component is dominated and at least a $4/5$ fraction of its vertices is uni-dominated: in the base cases by Theorem~\ref{thm:tree_full_cover}, and otherwise because the selected vertices uni-dominate every vertex of the component except possibly its root, which gives at least $n' - 1 \ge 4n'/5$ uni-dominated vertices for a component of order $n' \ge 6$. As the only neighbor of a side component outside it is $x$, and $x$ never belongs to the set chosen inside $Q$, these components require no later decision.

		Scanning the roots of the side components tells us whether $x$ already has a selected neighbor below it. If it does, the corresponding choice of Case 2 of Theorem~\ref{thm:tree_full_cover} for a vertex $x$ having a neighbor in $S_0$ is fixed at once. Otherwise the only neighbor of $x$ outside $Q$ whose status is still undecided is its parent $p$, and the bag $Q$ is attached to $p$: the first choice if $p \in S$, the second one if $p \notin S$. If $x = r$ there is no such $p$: the second choice is fixed at once, and the deletion of the subtree rooted at $x$ empties the active tree, so that only the propagation phase remains. In all cases $x \notin S_w$, so the bag does not dominate $p$. Every vertex of $Q$ is dominated and at least four of them are uni-dominated.

		\textbf{Terminal step.} When at most $5$ vertices are active, we solve the active tree by the corresponding base case of Theorem~\ref{thm:tree_full_cover}. When it has height at most $2$ and order at least $6$, we select its root if its height is at most $1$, and the children of its root if its height is $2$, as in the treatment of the heights $h \le 1$ and $h = 2$ in the induction step of Theorem~\ref{thm:tree_full_cover}. In both situations the active tree is dominated, at least $4/5$ of its vertices are uni-dominated, and all of them become decided.

		\textbf{Propagation.} Each bag is attached to a vertex that is still active when the bag is created, hence deleted strictly later. The dependencies are therefore directed towards the root and are acyclic. When the status of a vertex $z$ becomes known, every bag attached to $z$ is resolved, which fixes the status of its vertices and may in turn resolve further bags. The terminal step fixes the status of the last undecided vertices and triggers the resolution of all the remaining bags.

		\textbf{Correctness.} The bags, the side components and the terminal component partition $V(T)$. A bag of the case $k \ge 5$ contributes at least $k - 1 \ge 4k/5$ uni-dominated vertices out of $k$, a bag of the case $k = 4$ at least $4$ out of $5$, a bag of the case $k = 3$ exactly $3$ out of $3$, and each side component as well as the terminal component at least a $4/5$ fraction of its vertices. Summing gives $\ud(S) \ge 4n/5$, provided that no choice destroys the domination counted in another part. This is the content of Theorem~\ref{thm:tree_full_cover}, transposed as follows. In the case $k \ge 5$ the only vertex of the bag with a neighbor outside it is $w$, which is not counted, and no vertex of $S_w$ has a neighbor outside the bag. In the case $k = 3$ the only outside neighbor is $x$, and the rule stored in the bag distinguishes the two possible statuses of $x$. In the case $k = 4$ the only vertex of $Q$ with neighbors outside $Q$ is $x$, and the rule again distinguishes the two possible statuses of the outside neighbors of $x$.

		It remains to check that postponing the decision of a bag creates no conflict. Let $Q$ be a bag attached to a vertex $z$. Whichever of its two alternatives is eventually chosen, the vertex of $Q$ adjacent to $z$, namely $w$ in the case $k = 3$ and $x$ in the case $k = 4$, is not selected. Hence $Q$ contributes no selected neighbor to $z$, so the domination status of $z$, and of every vertex that is still active, does not depend on $Q$, and the active tree can be processed as if $Q$ had been removed. Conversely, when $Q$ is created, $z$ is the only vertex with a neighbor in $Q$ whose membership in $S$ is still undecided, and the rule stored in $Q$ prescribes a correct choice both for $z \in S$ and for $z \notin S$. Therefore, once all bags are resolved, every vertex of $T$ is dominated, and every vertex counted in a bag is uni-dominated by the final set $S$.

		\textbf{Running time.} The initialization takes $\mathcal{O}(n)$ time. To find an active vertex of maximum depth we scan the buckets $D_d$ from the largest index downwards, keeping a pointer made of the current bucket index and of a position inside that bucket. Neither component ever moves backwards, which is legitimate because the largest depth of an active vertex never increases and because a vertex passed over is deleted. Each vertex is thus skipped at most once, so all these searches cost $\mathcal{O}(n)$ in total. When $w$ is processed, its children and grandchildren are scanned in order to build $T_w$, and $T_w$ is then deleted. In the case $k = 4$ the children of $x$ are scanned once and $x$ is deleted, and each side component is traversed only when it is deleted. Every vertex and every edge is thus examined a constant number of times. Finally each bag is created once and resolved once, and each dependency is processed once. Since $|E(T)| = n - 1$, the total running time is $\mathcal{O}(n)$, and the algorithm uses $\mathcal{O}(n)$ space. \qed
	\end{proof}
	\subsection{Computational Complexity}\label{sec:fcud-complexity}
	We investigate the computational complexity of the Full-Cover Uni-domination (FCUD) problem, defined as follows.
	\begin{problembox}{Full-Cover Uni-domination (FCUD)}
		\textbf{Instance:} A graph $G = (V, E)$ and a positive integer $k$. \\
		\textbf{Question:} Is $\ud_0^*(G) \ge k$?
	\end{problembox}
	The complexity of FCUD depends on the threshold $k$. For a constant $k = c$, the dominators of $c$ uni-dominated vertices $C$ form a \emph{core} $S_C$ of at most $c$ vertices, and adding every vertex outside $N[C]$ keeps $C$ uni-dominated. It then suffices to test whether this expanded set dominates $G$, as formalized in Lemma~\ref{lem:canonical_set}.
	\begin{lemmarep}[$\star$] \label{lem:canonical_set}
		Let $G = (V,E)$ be a graph and $c \ge 1$ be an integer. For any subset $C \subseteq V$, let $R_C = V \setminus N[C]$ be the set of vertices that are neither in $C$ nor adjacent to a vertex of $C$.
		There exists a dominating set $S \subseteq V$ such that $\ud(S) \ge c$ if and only if there exist a target subset $C \subseteq V$ with $|C| = c$ and a core subset $S_C \subseteq V$ with $|S_C| \le c$ satisfying the following conditions:
		\begin{enumerate}
			\item $C \subseteq \UD(S_C)$.
			\item The expanded set $\widehat{S} = S_C \cup R_C$ is a dominating set of $G$.
		\end{enumerate}
	\end{lemmarep}
	\begin{proof}
		$(\Rightarrow)$ Assume there exists a dominating set $S$ of $G$ with $\ud(S) \ge c$. We can choose a target subset $C \subseteq \UD(S)$ of exactly $c$ uni-dominated vertices. Let $S_C = S \cap N[C]$. By definition, $N[C] = \bigcup_{u \in C} N[u]$, which allows us to distribute the intersection: $S_C = \bigcup_{u \in C} (S \cap N[u])$.
		Because each $u \in C$ is uni-dominated by $S$, the intersection $S \cap N[u]$ contains exactly one vertex. Since $N[u] \subseteq N[C]$, this unique dominator belongs to $S \cap N[C]$, which is exactly $S_C$. Thus, $S_C$ dominates $u$. Furthermore, because $S_C \subseteq S$ and $S$ uni-dominates $u$, $S_C$ cannot dominate $u$ more than once. Therefore, $S_C$ uni-dominates $u$, yielding $C \subseteq \UD(S_C)$. Additionally, since $S_C$ is the union of exactly $c$ singleton sets, its cardinality is bounded by $|S_C| \le c$.

		Next, we show that the expanded set $\widehat{S} = S_C \cup R_C$ is a dominating set of $G$. Consider an arbitrary vertex $v \in V$. We distinguish two cases:
		\begin{itemize}
			\item \textbf{Case 1:} $v \in R_C$. By the construction of $\widehat{S}$, $v$ is directly included in $\widehat{S}$, meaning it is dominated by itself.
			\item \textbf{Case 2:} $v \in N[C]$. Since the original set $S$ is a dominating set of $G$, $v$ must be dominated by some vertex $w \in S$. We consider where $w$ lies:
			\begin{itemize}
				\item \textbf{Subcase 2.1:} $w \in N[C]$. Because $w \in S$ and $w \in N[C]$, $w$ belongs to their intersection. By our earlier definition $S_C = S \cap N[C]$, which directly implies $w \in S_C$. Since $S_C \subseteq \widehat{S}$, the vertex $w$ is present in $\widehat{S}$.
				\item \textbf{Subcase 2.2:} $w \notin N[C]$. By the definition of the remaining set $R_C = V \setminus N[C]$, any vertex not in $N[C]$ must belong to $R_C$. Thus, $w \in R_C$. Since $R_C \subseteq \widehat{S}$, the vertex $w$ is again present in $\widehat{S}$.
			\end{itemize}
		\end{itemize}
		In all scenarios, the vertex responsible for dominating $v$ is guaranteed to be in $\widehat{S}$, proving that $\widehat{S}$ is indeed a dominating set.

		$(\Leftarrow)$ Assume such subsets $C$ and $S_C$ exist. By condition 2, $\widehat{S} = S_C \cup R_C$ is a dominating set of $G$. Consider an arbitrary vertex $u \in C$. Since $u \in C$, we have $N[u] \subseteq N[C]$, and therefore $N[u]$ is disjoint from $R_C = V \setminus N[C]$. Therefore, the intersection of its closed neighborhood with the expanded set $\widehat{S}$ simplifies to $N[u] \cap \widehat{S} = N[u] \cap (S_C \cup R_C) = N[u] \cap S_C$. By condition 1, $u$ is uni-dominated by $S_C$, meaning $|N[u] \cap S_C| = 1$. This implies $|N[u] \cap \widehat{S}| = 1$. Because this unique domination property holds for every single vertex $u \in C$, every such vertex is uni-dominated by $\widehat{S}$. Consequently, the entire target set $C$ is contained in $\UD(\widehat{S})$. Given that $|C| = c$, we conclude that $\ud(\widehat{S}) \ge c$. \qed
	\end{proof}
	Because this lemma reduces the search space to subsets of size bounded by $c$, we obtain a polynomial-time algorithm.
	\begin{theoremrep}[$\star$] \label{thm:poly_time}
		For any fixed integer $c \ge 2$, given a graph $G = (V, E)$, the problem of deciding whether $\ud_0^*(G) \ge c$ is solvable in polynomial time.
	\end{theoremrep}
	\begin{proof}
		Given a graph $G = (V,E)$ with $n = |V|$ and a fixed integer $c \ge 2$, we algorithmically verify if $\ud_0^*(G) \ge c$ by applying the expansion logic from Lemma~\ref{lem:canonical_set}:
		\begin{enumerate}
			\item Generate all possible target subsets $C \subseteq V$ of size exactly $c$ one by one.
			\item Generate all possible core subsets $S_C \subseteq V$ of size at most $c$ one by one.
			\item For each pair $(C, S_C)$, check if $S_C$ uni-dominates $C$. If it does, expand $S_C$ into $\widehat{S} = S_C \cup R_C$ (where $R_C = V \setminus N[C]$). Check if this expanded set $\widehat{S}$ dominates the entire graph $G$.
			\item If any pair successfully produces a global dominating set, return \texttt{TRUE}. If all combinations fail, return \texttt{FALSE}.
		\end{enumerate}
		To establish the time complexity, observe that since $c$ is fixed, the number of target subsets $C$ generated in Step 1 is $\binom{n}{c} = \mathcal{O}(n^c)$. Similarly, the number of core subsets $S_C$ generated in Step 2 is bounded by $\sum_{i=1}^c \binom{n}{i}$, which is also $\mathcal{O}(n^c)$. The algorithm therefore evaluates at most $\mathcal{O}(n^{2c})$ pairs of $(C, S_C)$. Inside the loop, for each pair, isolating the non-adjacent set $R_C$, verifying the local uni-domination on $C$, and confirming the global domination of $\widehat{S}$ can all be executed in $\mathcal{O}(n^2)$ time. The overall worst-case time complexity is bounded by $\mathcal{O}(n^{2c+2})$. This running time is polynomial, which concludes the proof. \qed
	\end{proof}
	In contrast, when the threshold grows with the order of the graph, the problem becomes NP-complete.
	\begin{theoremrep}[$\star$] \label{thm:np_complete}
		For any fixed integer $c \ge 2$, given a graph $G = (V, E)$, the problem of deciding whether $\ud_0^*(G) \ge \lceil \sqrt[c]{n} \rceil$ (where $n = |V|$) is NP-complete.
	\end{theoremrep}
	\begin{proofsketch}
		The problem is in NP, as for a given subset $S \subseteq V$ one can check in polynomial time whether $V_0(S) = \emptyset$ and $\ud(S) \ge \lceil \sqrt[c]{n} \rceil$. To prove NP-hardness, we perform a polynomial-time reduction from RX3C. The reduction is described below.
		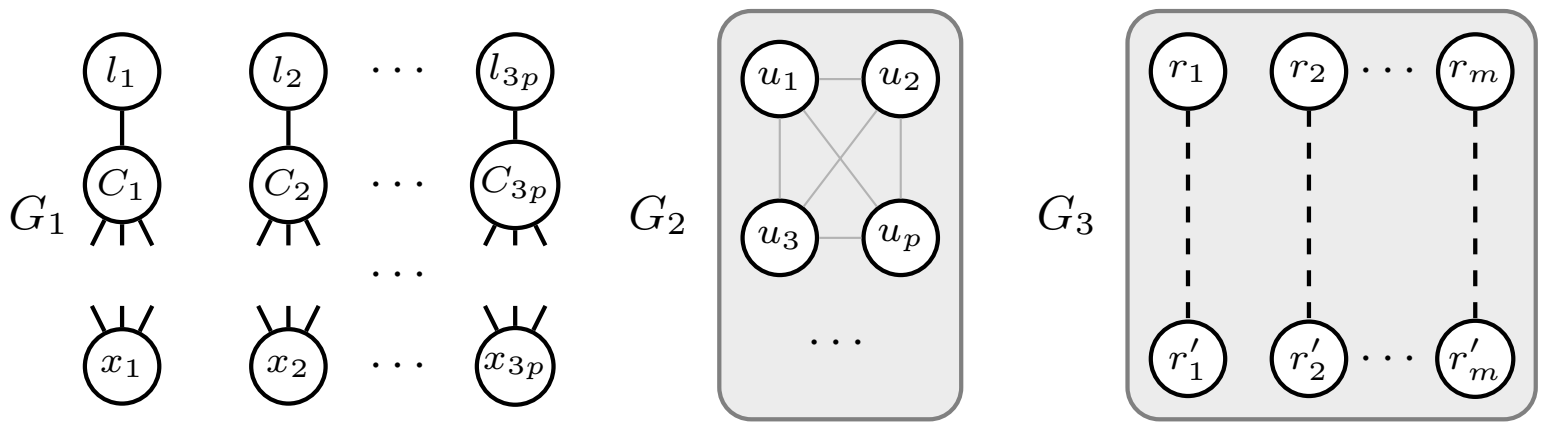
\begin{figure}[!ht]
			\centering
			\begin{tikzpicture}[
				vertex/.style={circle, draw, thick, fill=white, minimum size=14pt, inner sep=0.5pt, font=\scriptsize},
				dots/.style={inner sep=0pt, font=\small},
				norm edge/.style={thick},
				cross edge/.style={gray!60, thin},
				part/.style={fill=gray!15, rounded corners=6pt, thick, draw=black!50}
				]
				\node at (-0.55, 1.0) {$G_1$};
				\foreach \i/\x/\j in {1/0/1, 2/1.1/2, 3/2.6/{3p}} {
					\node[vertex] (l\i) at (\x, 1.95) {$l_{\j}$};
					\node[vertex] (c\i) at (\x, 1.2) {$C_{\j}$};
					\node[vertex] (x\i) at (\x, 0)   {$x_{\j}$};
					\draw[norm edge] (l\i) -- (c\i);
					\foreach \d in {-0.2, 0, 0.2} {
						\draw[norm edge] (c\i) -- ++(\d, -0.4);
						\draw[norm edge] (x\i) -- ++(\d, 0.4);
					}
				}
				\foreach \y in {1.95, 1.2, 0} \node[dots] at (1.85, \y) {$\cdots$};
				\node[dots] at (1.85, 0.6) {$\cdots$};
				\node at (3.55, 1.0) {$G_2$};
				\draw[part] (3.95, -0.35) rectangle (5.55, 2.35);
				\node[vertex] (u1) at (4.35, 1.9) {$u_1$};
				\node[vertex] (u2) at (5.15, 1.9) {$u_2$};
				\node[vertex] (u3) at (4.35, 0.85) {$u_3$};
				\node[vertex] (up) at (5.15, 0.85) {$u_p$};
				\draw[cross edge] (u1)--(u2) (u1)--(u3) (u1)--(up) (u2)--(u3) (u2)--(up) (u3)--(up);
				\node[dots] at (4.75, 0.15) {$\cdots$};
				\node at (6.25, 1.0) {$G_3$};
				\draw[part] (6.65, -0.35) rectangle (9.35, 2.35);
				\foreach \i/\x/\j in {1/7.05/1, 2/7.85/2, 3/8.95/m} {
					\node[vertex] (r\i) at (\x, 1.95) {$r_{\j}$};
					\node[vertex] (s\i) at (\x, 0.05) {$r'_{\j}$};
					\draw[dashed, thick] (r\i) -- (s\i);
				}
				\node[dots] at (8.4, 1.95) {$\cdots$};
				\node[dots] at (8.4, 0.05) {$\cdots$};
			\end{tikzpicture}
			\caption{The construction of graph $G$, the disjoint union of $G_1$, $G_2$ and $G_3$. Gray lines are edges and dashed lines are non-edges; the edges of $G_3$ are omitted.}
			\label{fig:hardness_construction_simplified}
		\end{figure}

		Let $I = (X, \mathcal{C})$ be an instance of RX3C, so that $|X| = |\mathcal{C}| = 3p$, and set $y = 10p + 2$. We construct $G$ as the disjoint union of three graphs $G_1$, $G_2$ and $G_3$. The part $G_1$ encodes the instance $I$, $G_2$ adds $p$ vertices so that ${|V(G_1)| + |V(G_2)| = 10p}$ is even, which makes the order of $G_3$ even, and $G_3$ is a padding part whose contribution to the uni-domination is a controlled constant and whose size brings the order of $G$ to a $c$-th power.
		\begin{itemize}
			\item Part $G_1$: We create three layers of vertices. The bottom layer consists of $3p$ vertices representing the elements $x \in X$. The middle layer consists of $3p$ vertices representing the subsets $C_j \in \mathcal{C}$. We add an edge between a vertex representing an element $x$ and a vertex representing a subset $C_j$ if and only if $x \in C_j$. Finally, the top layer consists of $3p$ pendant vertices $L = \{l_j : 1 \le j \le 3p\}$, the leaf $l_j$ being adjacent to the subset vertex $C_j$ only. Thus $|V(G_1)| = 9p$.
			\item Part $G_2$: We construct a complete graph $K_p$ on exactly $p$ vertices. Thus $|V(G_2)| = p$.
			\item Part $G_3$: Let $N_M = y^c - 10p$. Since $y = 10p + 2$ is even, $y^c$ is even, and as $10p$ is even, $N_M$ is an even integer. Moreover $y^c \ge (10p+2)^2 > 10p$, so $N_M \ge 2$. We let $G_3$ be a complete graph on $N_M$ vertices minus a perfect matching, that is, $V(G_3) = \{r_1, r'_1, \dots, r_m, r'_m\}$ where $m = N_M / 2$, all pairs of vertices being adjacent except the pairs $r_i r'_i$.
		\end{itemize}
		The construction is illustrated in Figure~\ref{fig:hardness_construction_simplified}. The total number of vertices of $G$ is $9p + p + (y^c - 10p) = y^c$, so the required threshold is $\lceil \sqrt[c]{n} \rceil = y = 10p + 2$. Since $c$ is fixed and the numbers of vertices and edges of $G$ are bounded by a polynomial in $p$, this instance is constructed in polynomial time in the size of $I$. It remains to prove that $I$ has an exact cover if and only if $G$ admits a dominating set $S$ with $\ud(S) \ge 10p + 2$ (see \fullver). \qed
	\end{proofsketch}
	\begin{proof}
		We use the notation of the reduction described in Section~\ref{sec:fcud-complexity}, and prove that the RX3C instance $I$ has an exact cover if and only if $G$ admits a dominating set $S$ with $\ud(S) \ge 10p + 2$.

		$(\Rightarrow)$ Suppose there exists an exact cover $\mathcal{C}^* \subseteq \mathcal{C}$, so that $|\mathcal{C}^*| = p$. In $G_1$, let $S_1 = \mathcal{C}^* \cup \{l_j : C_j \in \mathcal{C} \setminus \mathcal{C}^*\}$. In $G_2$, let $S_2$ consist of a single vertex. In $G_3$, let $S_3 = \{r_1, r'_1\}$, a pair of non-adjacent vertices. Set $S = S_1 \cup S_2 \cup S_3$.

		In $G_1$, every element of $X$ is adjacent to exactly one member of $\mathcal{C}^*$ and to no leaf, hence is uni-dominated. Every $C_j \in \mathcal{C}^*$ is dominated by itself only, since $l_j \notin S_1$ and no element vertex belongs to $S_1$. Every $C_j \notin \mathcal{C}^*$ is dominated by $l_j$ only. Every leaf $l_j$ is dominated by $C_j$ if $C_j \in \mathcal{C}^*$ and by itself otherwise. All $9p$ vertices of $G_1$ are therefore uni-dominated. In $G_2$, the chosen vertex is adjacent to all the others, so all $p$ vertices of $G_2$ are uni-dominated. In $G_3$, every vertex other than $r_1$ and $r'_1$ is adjacent to both of them and is dominated twice, whereas $r_1$ and $r'_1$ are dominated by themselves only. Hence $V_0(S) = \emptyset$ and $\ud(S) = 9p + p + 2 = 10p + 2$.

		$(\Leftarrow)$ Suppose there exists a dominating set $S$ of $G$ with $\ud(S) \ge 10p + 2$, and let $S_i = S \cap V(G_i)$. As $G$ is the disjoint union of $G_1$, $G_2$ and $G_3$, we have $\ud(S) = \ud(S_1) + \ud(S_2) + \ud(S_3)$, each term being computed inside the corresponding part. Trivially $\ud(S_1) \le 9p$, and $\ud(S_2) \le p$ because $G_2$ has $p$ vertices. In $G_3$ no vertex is universal, so $|S_3| \ge 2$. If $|S_3| = 2$, then exactly two vertices of $G_3$ are uni-dominated: the two chosen ones if they are non-adjacent, and their respective non-neighbors if they are adjacent. If $|S_3| \ge 3$, then applying Lemma~\ref{lem:non_neighbors} inside $G_3$, where every vertex has exactly one non-neighbor, shows that no vertex of $G_3$ is uni-dominated. In both cases $\ud(S_3) \le 2$.

		The three upper bounds add up to $9p + p + 2 = 10p + 2$, so each of them is attained. In particular every vertex of $G_1$ is uni-dominated.

		Suppose that some element vertex $x \in X$ belongs to $S_1$. To ensure that $x$ is uni-dominated, none of its neighbors can belong to $S_1$. Let $C_j$ be one of them. Since $C_j \notin S_1$ and $N[l_j] = \{l_j, C_j\}$, we must have $l_j \in S_1$. But then $C_j$ is dominated by both $x$ and $l_j$, contradicting the fact that $C_j$ is uni-dominated. Hence $S_1 \cap X = \emptyset$.

		Consequently the elements of $X$ are dominated only by vertices of $S_1 \cap \mathcal{C}$, and since all elements of $X$ are uni-dominated, the subsets of $S_1 \cap \mathcal{C}$ cover every element of $X$ exactly once. They form an exact cover of $X$. \qed
	\end{proof}
	\section{The Non-Overlapping Uni-domination Problem}\label{sec:no-complexity}
	Throughout this section $V_+(S) = \emptyset$. Every vertex it dominates is then dominated exactly once, so $\ud(S) = I(S) = \sum_{v \in S}(1 + \deg(v))$.
	\subsection{Structural Properties}
	Unlike the full-cover variant, which may be bounded by a constant, the non-overlapping constraint yields a number of uni-dominated vertices that always scales with the order of the graph. Goddard, Oellermann, Slater and Swart \cite{goddard2000bounds} proved the sharp bound $\ud_+^*(G) \ge 1 + \sqrt{n-1}$ for every graph of order $n$, and exhibited graphs of diameter $2$ obtained from projective planes for which $\ud_+^*(G) = \lceil 1 + \sqrt{n-1} \rceil$.

	We record one observation about that bound. Their argument starts from a $2$\nobreakdash-packing of maximum influence, but the proof of the two inequalities it combines, $\ud_+^*(G) \ge 1 + \Delta$ and $\ud_+^*(G) \ge n(\Delta+1)/(\Delta^2+1)$, uses only that the $2$-packing under consideration contains a vertex of maximum degree and is \emph{maximal}, respectively: a maximal $2$-packing $S$ satisfies $N[N[S]] = V$, which is all that the counting argument requires. A maximal $2$-packing containing a vertex of maximum degree is obtained in polynomial time by starting from $S = \{v\}$ with $\deg(v) = \Delta$ and, as long as some vertex is at distance at least $3$ from every vertex of $S$, adding such a vertex to $S$. As the first bound is increasing and the second one is decreasing in $\Delta \ge 1$, and the two coincide at $\Delta = \sqrt{n-1}$, their maximum is at least $1 + \sqrt{n-1}$ for every $\Delta$. Hence such a set $S$ already satisfies $\ud(S) \ge 1 + \sqrt{n-1}$, so the bound of \cite{goddard2000bounds} is attained constructively.

	No lower bound linear in $n$ holds, even on trees: the same authors proved the sharp bound $\ud_+^*(T) \ge \sqrt{8(n+2)} - 4$ for every tree of order $n$, together with an extremal family attaining it \cite{goddard2000bounds}. The contrast with Theorem~\ref{thm:tree_full_cover}, which guarantees $4n/5$ uni-dominated vertices on every tree under full coverage, is the sharpest illustration of how differently the two constraints behave.
	\subsection{Computational Complexity}
	We consider the following decision problem.
	\begin{problembox}{Non-Overlapping Uni-domination}
		\textbf{Instance:} A graph $G = (V, E)$ and a positive integer $k$. \\
		\textbf{Question:} Is $\ud_+^*(G) \ge k$?
	\end{problembox}
	For a fixed constant threshold $k = c$, the answer is positive whenever $n \ge (c-1)^2 + 1$, by the bound $\ud_+^*(G) \ge 1 + \sqrt{n-1}$ recalled above, and the remaining instances have bounded order, so they are settled by exhaustive search in constant time. We show that raising the threshold by a constant factor, from $1 + \sqrt{n-1}$ to $4\sqrt{n}$, already makes the problem NP-complete on split graphs.
	\begin{theoremrep}[$\star$] \label{thm:np_complete_sqrt}
		Given a graph $G$ with $n$ vertices, the problem of deciding whether $\ud_+^*(G) \ge 4\sqrt{n}$ is NP-complete, even for split graphs.
	\end{theoremrep}
	\begin{proofsketch}
		The problem is in NP, as verifying whether a given subset $S \subseteq V$ satisfies $V_+(S) = \emptyset$ and $\ud(S) \ge 4\sqrt{n}$ can be done in polynomial time. To prove NP-hardness, we present a polynomial-time reduction from the RX3C problem. Let $I = (X, \mathcal{C})$ be an instance of RX3C, with $|X| = 3p$ and $\mathcal{C} = \{C_1, \dots, C_{3p}\}$. Instances with $p \le 7$ have bounded size and are solved in constant time, so we may assume $p \ge 8$.
		\begin{figure}[!ht]
			\centering
			\begin{tikzpicture}[
				vertex/.style={circle, draw, thick, fill=white, minimum size=16pt, inner sep=0pt, font=\scriptsize},
				dots/.style={inner sep=0pt, font=\small},
				norm edge/.style={thick}
				]
				\draw[fill=gray!15, rounded corners=6pt, thick, draw=black!50] (-0.9, -0.36) rectangle (10.3, 0.36);
				\node at (-0.55, 0) {$X$};
				\node at (-0.55, 1.5) {$\mathcal{C}$};
				\node[vertex] (c11) at (0.3, 1.5) {$v_1^1$};
				\node[dots] at (1.0, 1.5) {$\cdots$};
				\node[vertex] (c1x) at (1.7, 1.5) {$v_1^{s_1}$};
				\draw[decorate,decoration={brace,amplitude=3pt}, thick] (0.0, 1.9) -- (2.0, 1.9) node[midway,above=2pt,font=\scriptsize] {$s_1$ vertices};
				\node[vertex] (ck1) at (8.2, 1.5) {$v_{3p}^1$};
				\node[dots] at (8.95, 1.5) {$\cdots$};
				\node[vertex] (ckx) at (9.7, 1.5) {\scalebox{0.85}{$v_{3p}^{s_{3p}}$}};
				\draw[decorate,decoration={brace,amplitude=3pt}, thick] (7.85, 1.9) -- (10.05, 1.9) node[midway,above=2pt,font=\scriptsize] {$s_{3p}$ vertices};
				\node[dots] at (5.0, 1.5) {$\cdots$};
				\node[vertex] (x1) at (0.3, 0) {$x_1$};
				\node[vertex] (x2) at (1.7, 0) {$x_2$};
				\node[dots] at (5.0, 0) {$\cdots$};
				\node[vertex] (xk1) at (8.2, 0) {\scalebox{0.8}{$x_{3p-1}$}};
				\node[vertex] (xk) at (9.7, 0) {$x_{3p}$};
				\foreach \n in {c11, c1x, ck1, ckx} \foreach \d in {-0.2,0,0.2} \draw[norm edge] (\n) -- ++(\d, -0.45);
				\foreach \n in {x1, x2, xk1, xk}  \foreach \d in {-0.15,0,0.15} \draw[norm edge] (\n) -- ++(\d, 0.45);
				\node[dots] at (5.0, 0.75) {$\cdots$};
			\end{tikzpicture}
			\caption{The graph $G$ built from an RX3C instance in the proof of Theorem~\ref{thm:np_complete_sqrt}. The shaded rectangle is the clique $X$, whose edges are omitted, and dots stand for omitted vertices.}
			\label{fig:hardness_construction_sqrt}
		\end{figure}
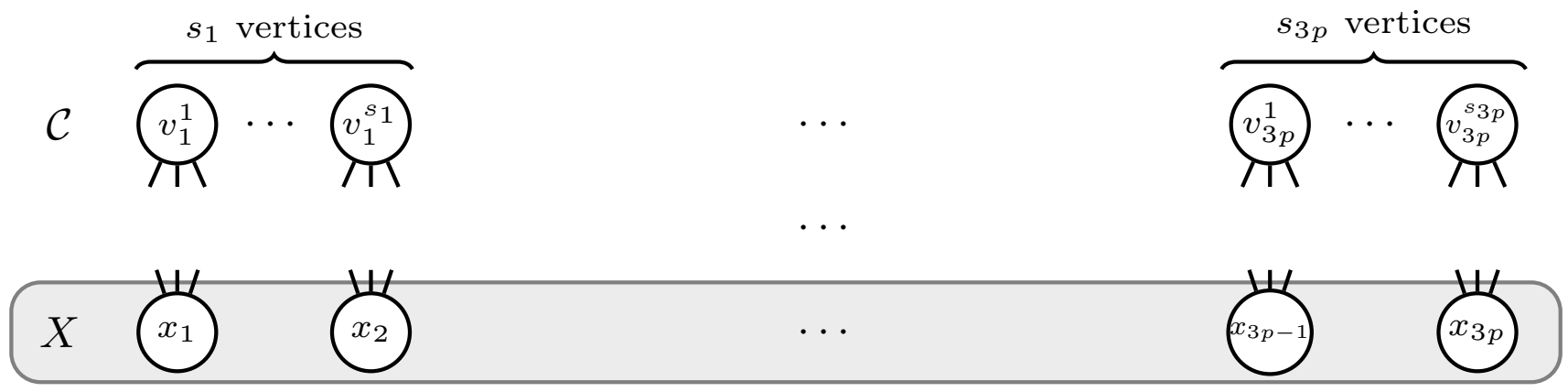

		We construct $G = (V,E)$ as follows. Each element of $X$ becomes a vertex, and $X$ is turned into a clique of size $3p$. Each subset $C_i \in \mathcal{C}$ receives a gadget consisting of an independent set of $s_i$ vertices, each of them joined to the three vertices of $X$ corresponding to the elements of $C_i$. Since $V$ is partitioned into the clique $X$ and the independent set of all gadget vertices, $G$ is a split graph.

		The construction is illustrated in Figure~\ref{fig:hardness_construction_sqrt}, and the sizes $s_i$ are chosen so that the order $n$ of $G$ equals $p^2$. Let $q = \lfloor p/3 \rfloor$, and note that $q \ge 2$ because $p \ge 8$, so every gadget receives at least one vertex. If $p = 3q$, set $s_i = q-1$ for all $3p$ gadgets, so that $n = 3p + 3p(q-1) = 3pq = p^2$. If $p = 3q+1$, set $s_i = q$ for exactly $p$ gadgets and $s_i = q-1$ for the other $2p$, so that $n = 3p + pq + 2p(q-1) = p + 3pq = p^2$. If $p = 3q+2$, set $s_i = q$ for exactly $2p$ gadgets and $s_i = q-1$ for the other $p$, so that $n = 3p + 2pq + p(q-1) = 2p + 3pq = p^2$. In all three cases $n = p^2$, so $p = \sqrt{n}$ and the required threshold is $4p = 4\sqrt{n}$. So $G$ has $p^2$ vertices and $\mathcal{O}(p^2)$ edges, hence is built in polynomial time. It remains to prove that $I$ has an exact cover if and only if $G$ admits a $2$-packing $S$ with $\ud(S) \ge 4p$ (see \fullver). \qed
	\end{proofsketch}
	\begin{proof}
		We use the notation of the reduction described in Section~\ref{sec:no-complexity}, and prove that the RX3C instance $I$ has an exact cover if and only if $G$ admits a $2$-packing $S$ such that $\ud(S) \ge 4p$.

		$(\Rightarrow)$ Suppose $I$ has an exact cover consisting of exactly $p$ subsets $C_{i_1}, \dots, C_{i_p}$. Let $S$ consist of exactly one vertex from the gadget of each selected subset, which is possible since every gadget is non-empty, yielding $|S| = p$. Since the chosen subsets form an exact cover of $X$, their neighborhoods in $X$ are pairwise disjoint. Because the chosen vertices also belong to independent sets, we have $V_+(S) = \emptyset$. Each of the $p$ vertices in $S$ dominates itself and exactly 3 vertices in $X$. This yields $\ud(S) = \sum_{v \in S} (1 + 3) = 4|S| = 4p = 4\sqrt{n}$, satisfying the threshold.

		$(\Leftarrow)$ Conversely, assume there exists a set $S \subseteq V$ with $V_+(S) = \emptyset$ and $\ud(S) \ge 4p$. We analyze the intersection of $S$ with the clique $X$:
		\begin{itemize}
			\item \textbf{Case 1:} $|S \cap X| \ge 2$. Since $X$ is a clique, two vertices of $S \cap X$ are adjacent, so each of them is dominated twice, contradicting $V_+(S) = \emptyset$.
			\item \textbf{Case 2:} $|S \cap X| = 1$. Let $S \cap X = \{u\}$. If $S$ contained a gadget vertex $v$, then $v$ would dominate its 3 neighbors in $X$. Since $u$ lies in the clique $X$, it also dominates these 3 vertices. Those 3 vertices of $X$ are then over-dominated, forcing $S$ to be a singleton ($S = \{u\}$).
			The vertex $u$ dominates itself, the remaining $3p - 1$ vertices of $X$, and the vertices of the three gadgets whose subsets contain $u$. Let these subsets containing $u$ be $C_i, C_j$, and $C_k$. The number of uni-dominated vertices is bounded by $\ud(S) \le 1 + (3p - 1) + s_i + s_j + s_k = 3p + s_i + s_j + s_k$.
			We bound $s_i + s_j + s_k$ separately in each congruence class:
			\begin{itemize}
				\item If $p \equiv 0 \pmod 3$, every gadget has size $q - 1 = \dfrac{p}{3} - 1$, so $\ud(S) \le 3p + (p - 3) = 4p - 3 < 4p$.
				\item If $p \equiv 1 \pmod 3$, every gadget has size at most $q = \dfrac{p - 1}{3}$, so $\ud(S) \le 3p + (p - 1) = 4p - 1 < 4p$.
				\item If $p \equiv 2 \pmod 3$, every gadget has size at most $q = \dfrac{p - 2}{3}$, so $\ud(S) \le 3p + (p - 2) = 4p - 2 < 4p$.
			\end{itemize}
			In all cases, the configuration fails to reach the required threshold of $4p$.
			\item \textbf{Case 3:} $|S \cap X| = 0$. This forces $S$ to consist entirely of gadget vertices. There can be at most one vertex chosen from each gadget, otherwise, two vertices from the same gadget would share their three neighbors in $X$, contradicting $V_+(S) = \emptyset$.

			By assumption $\ud(S) \ge 4\sqrt{n} = 4p$. Since each vertex in $S$ uni-dominates exactly four vertices (itself and three elements in $X$), we must have $4|S| \ge 4p$, yielding $|S| \ge p$.

			Furthermore, to maintain $V_+(S) = \emptyset$, the corresponding subsets in $X$ must be pairwise disjoint. Since $X$ has $3p$ elements and each selected gadget covers three of them, we can select at most $p$ gadgets ($3|S| \le 3p$).

			Consequently, we must have exactly $|S| = p$. Because these $p$ disjoint subsets cover all $3p$ elements of $X$, they constitute an exact cover for the RX3C instance.
		\end{itemize}
		Since only Case 3 meets the threshold, $G$ admits a $2$-packing $S$ with $\ud(S) \ge 4\sqrt{n}$ if and only if the RX3C instance has a solution. \qed
	\end{proof}

	\section{Comparison of the Two Variants}\label{sec:compare}
	We now show that neither parameter bounds the other, by exhibiting a family of graphs for each direction.

	The first one consists of the complete graphs of even order $n \ge 4$ minus a perfect matching, on which $\ud_+^*$ exceeds $\ud_0^*$ by a factor of order $n$.
	\begin{theoremrep}[$\star$] \label{thm:separation_1}
		For every even integer $n \ge 4$, there exists a graph $G = (V,E)$ of order $n$ such that $2 = \ud_0^*(G) < \ud_+^*(G) = n - 1$.
	\end{theoremrep}
	\begin{proof}
		Let $n \ge 4$ be an even integer and let $G$ be the complete graph on $n$ vertices minus a perfect matching, so that each vertex is adjacent to all the others except exactly one non-neighbor. This is precisely the graph constructed in the proof of Theorem~\ref{thm:general_tightness} with $c = 2$ and $p = (n-2)/2$: the independent set $\mathcal{C} = \{x_1, x_2\}$ is the pair of ends of one non-edge, and the non-edges inside $R$ form the perfect matching of the remaining $2p$ vertices. Hence $\ud_0^*(G) = 2$ by Theorem~\ref{thm:general_tightness}. We now evaluate $\ud_+^*(G)$.

		Let $S$ be a set with $V_+(S) = \emptyset$ and suppose $|S| \ge 2$, say $u, v \in S$ with $u \neq v$. If $uv \in E$, then $u$ is dominated by both $u$ and $v$, a contradiction. If $uv \notin E$, then $u$ and $v$ are the two ends of a non-edge of the perfect matching, so the non-neighbor of any vertex $w \notin \{u,v\}$ is neither $u$ nor $v$. Since $n \ge 4$ such a $w$ exists, and it is adjacent to both $u$ and $v$, hence over-dominated, again a contradiction. Hence $|S| \le 1$.

		For a singleton $S = \{v\}$, the vertex $v$ uni-dominates itself and all its $n - 2$ neighbors. No vertex is over-dominated, and only its non-neighbor remains undominated. This configuration directly yields $\ud_+^*(G) = n - 1$.

		Hence, for every even integer $n \ge 4$, we have $2 = \ud_0^*(G) < n - 1 = \ud_+^*(G)$. \qed
	\end{proof}
	The second family consists of trees. For an integer $p \ge 3$, we construct a tree $T_p = (V,E)$ as follows: create a root vertex $r$; create $p$ star graphs $A_1, A_2, \dots, A_p$, where each star $A_i$ consists of a center vertex $r_i$ and $p-1$ leaves; add an edge between the root $r$ and each center $r_i$. We set $\mathcal{T} = \{T_p : p \ge 3\}$. The restriction $p \ge 3$ is necessary: the same construction with $p = 2$ produces the path $P_5$, which admits an efficient dominating set, so that both parameters equal $n$ and no separation occurs. The following theorem shows that on these trees the full-cover parameter exceeds the non-overlapping one by a factor of order $\sqrt{n}$.
	\begin{theoremrep}[$\star$] \label{thm:separation_2}
		Every tree $T_p \in \mathcal{T}$ has order $n = p^2 + 1$ and satisfies $3\sqrt{n-1} - 1 = \ud_+^*(T_p) < \ud_0^*(T_p) = n - 1$.
	\end{theoremrep}
	\begin{proof}
		Let $T_p \in \mathcal{T}$, so that $p \ge 3$. The construction is illustrated in Figure~\ref{fig:Tp_tree}. The number of vertices is $|V| = n = 1 + p \cdot (1 + p - 1) = p^2 + 1$, which implies $p = \sqrt{n-1}$.

		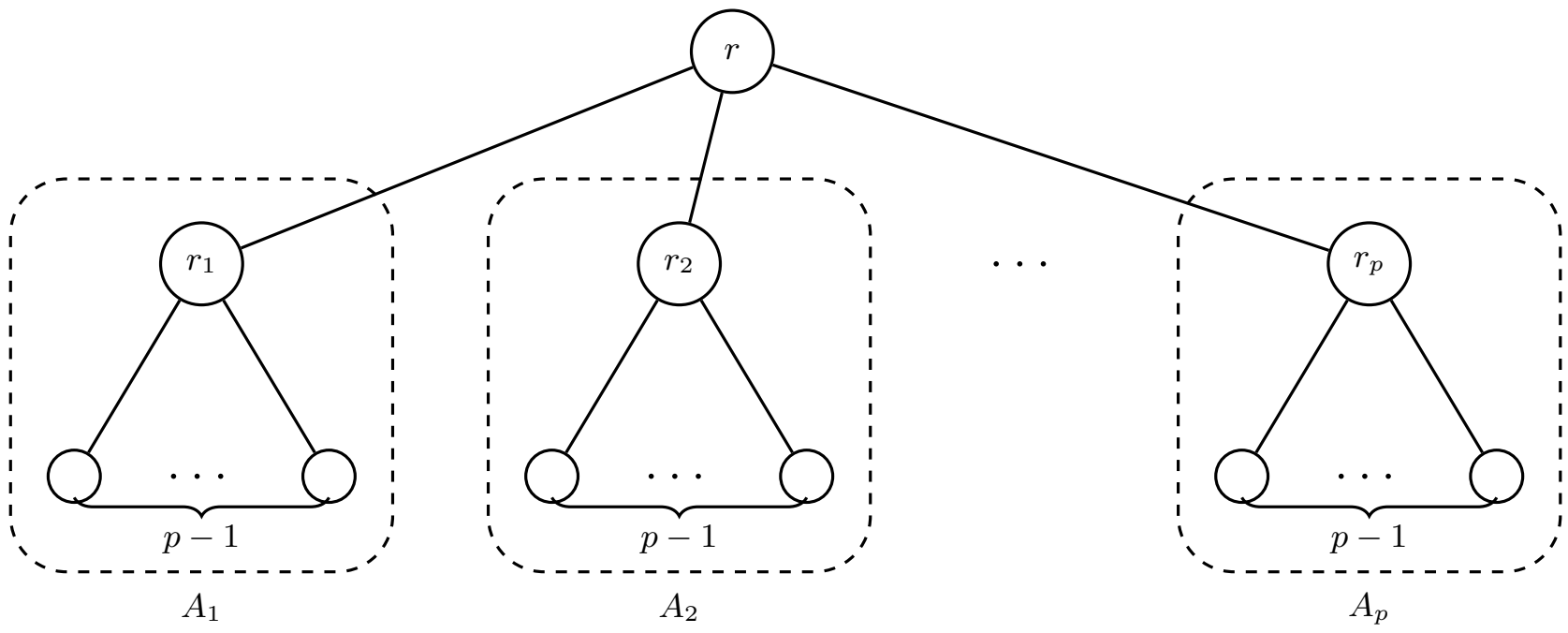
\begin{figure}[H]
			\centering
			\scalebox{0.75}{
				\begin{tikzpicture}[
					vertex/.style={circle, draw, thick, minimum size=22pt, inner sep=0pt},
					leaf/.style={circle, draw, thick, minimum size=14pt, inner sep=0pt},
					dots/.style={circle, draw=none, minimum size=0pt, inner sep=2pt},
					edge/.style={thick}
					]
					\node[vertex] (r) at (0,0) {$r$};
					\node[vertex] (r1) at (-5,-2) {$r_1$};
					\node[vertex] (r2) at (-0.5,-2) {$r_2$};
					\node[dots] (rdots) at (2.75,-2) {\Large $\dots$};
					\node[vertex] (rp) at (6,-2) {$r_p$};
					\draw[edge] (r) -- (r1);
					\draw[edge] (r) -- (r2);
					\draw[edge] (r) -- (rp);
					\node[leaf] (l11) at (-6.2,-4) {};
					\node[dots] (l1dots) at (-5,-4) {\Large $\dots$};
					\node[leaf] (l1p) at (-3.8,-4) {};
					\draw[edge] (r1) -- (l11);
					\draw[edge] (r1) -- (l1p);
					\node[leaf] (l21) at (-1.7,-4) {};
					\node[dots] (l2dots) at (-0.5,-4) {\Large $\dots$};
					\node[leaf] (l2p) at (0.7,-4) {};
					\draw[edge] (r2) -- (l21);
					\draw[edge] (r2) -- (l2p);
					\node[leaf] (lp1) at (4.8,-4) {};
					\node[dots] (lpdots) at (6,-4) {\Large $\dots$};
					\node[leaf] (lpp) at (7.2,-4) {};
					\draw[edge] (rp) -- (lp1);
					\draw[edge] (rp) -- (lpp);
					\draw[decorate,decoration={brace,amplitude=5pt,mirror}, thick]
					(-6.2,-4.2) -- (-3.8,-4.2) node[midway,below=4pt] {\small $p-1$};
					\draw[decorate,decoration={brace,amplitude=5pt,mirror}, thick]
					(-1.7,-4.2) -- (0.7,-4.2) node[midway,below=4pt] {\small $p-1$};
					\draw[decorate,decoration={brace,amplitude=5pt,mirror}, thick]
					(4.8,-4.2) -- (7.2,-4.2) node[midway,below=4pt] {\small $p-1$};
					\draw[dashed, thick, rounded corners=15pt] (-6.8,-4.9) rectangle (-3.2,-1.2);
					\node[dots, below] at (-5,-4.9) {$A_1$};
					\draw[dashed, thick, rounded corners=15pt] (-2.3,-4.9) rectangle (1.3,-1.2);
					\node[dots, below] at (-0.5,-4.9) {$A_2$};
					\draw[dashed, thick, rounded corners=15pt] (4.2,-4.9) rectangle (7.8,-1.2);
					\node[dots, below] at (6,-4.9) {$A_p$};
				\end{tikzpicture}
			}
			\caption{The structure of the tree $T_p$.}
			\label{fig:Tp_tree}
		\end{figure}
		We first evaluate $\ud_+^*(T_p)$ by considering any subset $S \subseteq V$ satisfying $V_+(S) = \emptyset$, that is, any $2$-packing $S$.

		\noindent \textbf{Case 1:} $r \in S$.

		The maximum distance from $r$ to any vertex in $T_p$ is 2. Since any two distinct vertices in $S$ must be separated by a distance of at least 3, $S$ can contain no other vertices. Thus, $S = \{r\}$, yielding:
		$$ \ud(S) = |N[r]| = 1 + p $$
		\textbf{Case 2:} $r \notin S$ and $S \cap \{r_1, \dots, r_p\} \neq \emptyset$.

		Assume without loss of generality that $r_1 \in S$. Because $d(r_1, r_i) = 2$ for all $i \neq 1$, $S$ cannot contain any other center $r_i$. Thus, $S \cap \{r_1, \dots, r_p\} = \{r_1\}$. The vertex $r_1$ dominates itself, the root $r$, and its $p-1$ leaves, yielding a total of $1 + 1 + (p-1) = p+1$ vertices.

		Any other vertex in $S$ must be a leaf in some star $A_i$ for $i \neq 1$. In any star $A_i$, the distance between any two leaves is 2 (via $r_i$). Hence $S$ contains at most one leaf from each $A_i$. Each chosen leaf dominates itself and its center $r_i$, contributing exactly 2 vertices. Across the remaining $p-1$ stars, these leaves contribute at most $2(p-1)$ vertices. In total, this case dominates at most:
		$$ \ud(S) \le (p + 1) + 2(p - 1) = 3p - 1 $$
		\textbf{Case 3:} $r \notin S$ and $S \cap \{r_1, \dots, r_p\} = \emptyset$.

		All vertices in $S$ must be leaves. As established in Case 2, $S$ can contain at most one leaf per star. With $p$ stars available, $S$ contains at most $p$ leaves. Each leaf dominates two vertices (itself and its center). The total dominated vertices is bounded by:
		$$ \ud(S) \le 2p $$
		Comparing the three cases, and using $p \ge 3$, we obtain, for every $2$-packing $S$,
		$$ \ud(S) \le \max\{p+1,\ 3p-1,\ 2p\} = 3p - 1. $$
		This value is attained. Let $S^\star$ consist of the center $r_1$ together with one leaf of $A_i$ for each $i \neq 1$. Then $S^\star$ is a $2$-packing: a leaf of $A_i$ is at distance $3$ from $r_1$, and two leaves belonging to two distinct stars are at distance $4$. Since $\ud(S) = \sum_{v \in S}(1 + \deg(v))$ for every $2$-packing $S$, we get $\ud(S^\star) = (1 + p) + 2(p-1) = 3p - 1$, so $\ud_+^*(T_p) = 3p - 1$.

		We now evaluate $\ud_0^*(T_p)$. Consider the subset $S$ consisting of all center vertices, $S = \{r_1, r_2, \dots, r_p\}$. Each center $r_i$ dominates itself, its $p-1$ leaves and the root $r$.

		Thus, every leaf is dominated exactly once (by its respective center $r_i$). Every center $r_i$ is dominated exactly once (by itself, since the distance between any two distinct centers is 2, so they do not dominate each other). The root $r$ is adjacent to all $p$ centers, so it is dominated $p$ times.

		Since $p \ge 3$, the root $r$ is over-dominated ($r \in V_+(S)$). All other $n-1$ vertices (leaves and centers) are uni-dominated, and $V_0(S) = \emptyset$.

		This configuration yields $\ud(S) = n - 1$, so $\ud_0^*(T_p) \ge n-1$.

		Conversely, $T_p$ has no dominating set uni-dominating all of its $n$ vertices, that is, no efficient dominating set. Indeed, let $S'$ be such a set and consider a star $A_i$. If $r_i \notin S'$, then every leaf of $A_i$ must belong to $S'$, since a leaf has $r_i$ as its only neighbor. As $A_i$ has $p - 1 \ge 2$ leaves, the center $r_i$ would then be dominated at least twice, a contradiction. Hence $r_i \in S'$ for every $i$, and the root $r$ is dominated $p \ge 3$ times, again a contradiction. Therefore $\ud_0^*(T_p) = n - 1$.

		Comparing the two parameters for any $p \ge 3$, we have $\ud_+^*(T_p) = 3p - 1 < p^2 = n - 1 = \ud_0^*(T_p)$, which completes the proof. \qed
	\end{proof}
	\section{Conclusion and Open Problems}
	We studied two restricted variants of the uni-domination problem, which measure how closely a graph can approach efficient domination when one of the two possible defects is forbidden. Besides the linear-time bound $\ud_0^*(T) \ge 4n/5$ on trees, which answers Problem~1 of \cite{HedetniemiRall2026}, we showed that the two variants differ sharply, both in their lower bounds and in the thresholds at which their decision problems become NP-complete, and that neither of them bounds the other.

	A quantitative gap remains for the non-overlapping variant: every graph satisfies $\ud_+^*(G) \ge 1 + \sqrt{n-1}$, and this is sharp, yet deciding whether $\ud_+^*(G) \ge 4\sqrt{n}$ is NP-complete, so that the complexity of the intermediate thresholds $c\sqrt{n}$ with $1 < c < 4$ is unknown. It would be interesting to tighten this gap, and to study these parameters, algorithmically and structurally, on graph classes such as planar or chordal graphs and their subclasses.

	\begin{credits}
		\subsubsection{\ackname} This work was supported by the International Research Center ``Innovation Transportation and Production Systems'' of the I-SITE CAP 20-25.
	\end{credits}
	\bibliographystyle{splncs04}
	\bibliography{references}

\end{document}